\documentclass[12pt]{elsart}

\usepackage{graphicx}
\usepackage{subfigure}
\usepackage{amsmath}
\usepackage{cite}
\usepackage{url}
\usepackage{endfloat}

\journal{Journal of Magnetism and Magnetic Materials}

\begin{document}

\begin{frontmatter}

\title{Effects of selective dilution on phase diagram and ground-state magnetizations of an Ising antiferromagnet on triangular and honeycomb lattices}
\author{M. \v{Z}ukovi\v{c}\corauthref{cor}},
\ead{milan.zukovic@upjs.sk}
\author{M. Borovsk\'{y}},
\ead{borovsky.michal@gmail.com}
\author{A. Bob\'ak}
\ead{andrej.bobak@upjs.sk}
\address{Department of Theoretical Physics and Astrophysics, Faculty of Science,\\ 
P. J. \v{S}af\'arik University, Park Angelinum 9, 041 54 Ko\v{s}ice, Slovak Republic}
\corauth[cor]{Corresponding author.}

\begin{abstract}
We employ an effective-field theory with correlations in order to study the phase diagram and ground-state magnetizations of a selectively diluted Ising antiferromagnet on triangular and honeycomb lattices. Dilution of different sublattices with generally unequal probabilities results in a rather intricate phase diagram in the sublattice dilution parameters space. In the case of the frustrated triangular lattice antiferromagnet the selective dilution affects the degree of frustration which can lead to some peculiar phenomena, such as reentrant behavior of long-range order or unsaturated sublattice magnetizations at zero temperature. The selectively diluted Ising antiferromagnet on the honeycomb lattice is obtained as a special case when one sublattice of the triangular lattice is completely removed by dilution.
\end{abstract}


\end{frontmatter}

\section{Introduction}
It is known that the pure Ising model with nearest-neighbor antiferromagnetic interactions on a triangular lattice shows no long-range ordering due to a high degree of frustration present~\cite{wannier,hout}. However, the typical high degeneracy of the ground state can be lifted by a small perturbation, such as an external magnetic field or the presence of quenched vacancies. The former case has been shown for a certain range of the field values to result in a phase transition between the ferrimagnetic phase with two sublattices aligned parallel and one antiparallel to the field ($\downarrow\uparrow\uparrow$) at lower temperatures and the paramagnetic phase in which all spins are aligned parallel to the field ($\uparrow\uparrow\uparrow$) at higher temperatures~\cite{metcalf,schick,netz}. A characteristic feature of the ferrimagnetic phase is a broad $m=1/3$ magnetization plateau. On the other hand, introduction of quenched magnetic vacancies in zero field locally relieves frustration and supposedly leads to a spin-glass order~\cite{grest,ander,blac}. A recent Monte Carlo study of magnetization processes in the system with both the field and uniform magnetic dilution showed that, for example, even a small amount of vacancies can deform the broad frustration-induced $1/3$ magnetization plateau into a stepwise curve~\cite{yao}. In our recent study of the critical behavior of such a system we found that the interplay between the applied field and the frustration-relieving dilution results in peculiar phase diagrams in the temperature-field-dilution parameter space, involving multiple reentrants~\cite{zuko}. Even more interesting is the case when, instead of uniform diluting, nonmagnetic vacancies are introduced into the lattice selectively. Namely, if the lattice is split into several (in the case of the triangular lattice three) sublattices which are populated by the vacancies with different probabilities, then the geometrical frustration can also be relieved globally, giving rise to long-range magnetic ordering phenomena even in the absence of the field. Kaya and Berker~\cite{kaya} have shown that if only one out of three sublattices of the triangular lattice is randomly diluted then a long-range order can develop in the remaining two sublattices already at relatively low concentrations of the vacancies, while the diluted sublattice remains disordered at any concentration of the vacancies. Also Monte Carlo simulations~\cite{robi} have supported this finding. \\
\hspace*{5mm} The goal of the present study is to extend the study of Kaya and Berker~\cite{kaya} in order to establish a three-dimensional phase diagram of the system with two sublattices jointly diluted but with different probabilities. Since the concentrations of the magnetic sites of the diluted sublattices represent two independent parameters that are able to control the degree of the frustration of the entire system, we can expect a nontrivial behavior displaying qualitative differences in different regions of the parameters space. Moreover, if one sublattice is completely diluted, i.e., effectively removed, the lattice reduces to the honeycomb type. Thus, by diluting the second sublattice we can obtain the phase diagram of the selectively diluted Ising antiferromagnet on the honeycomb lattice as a special case. 

\section{Formalism}
\begin{figure}[t!]
\centering
    \includegraphics[scale=.7]{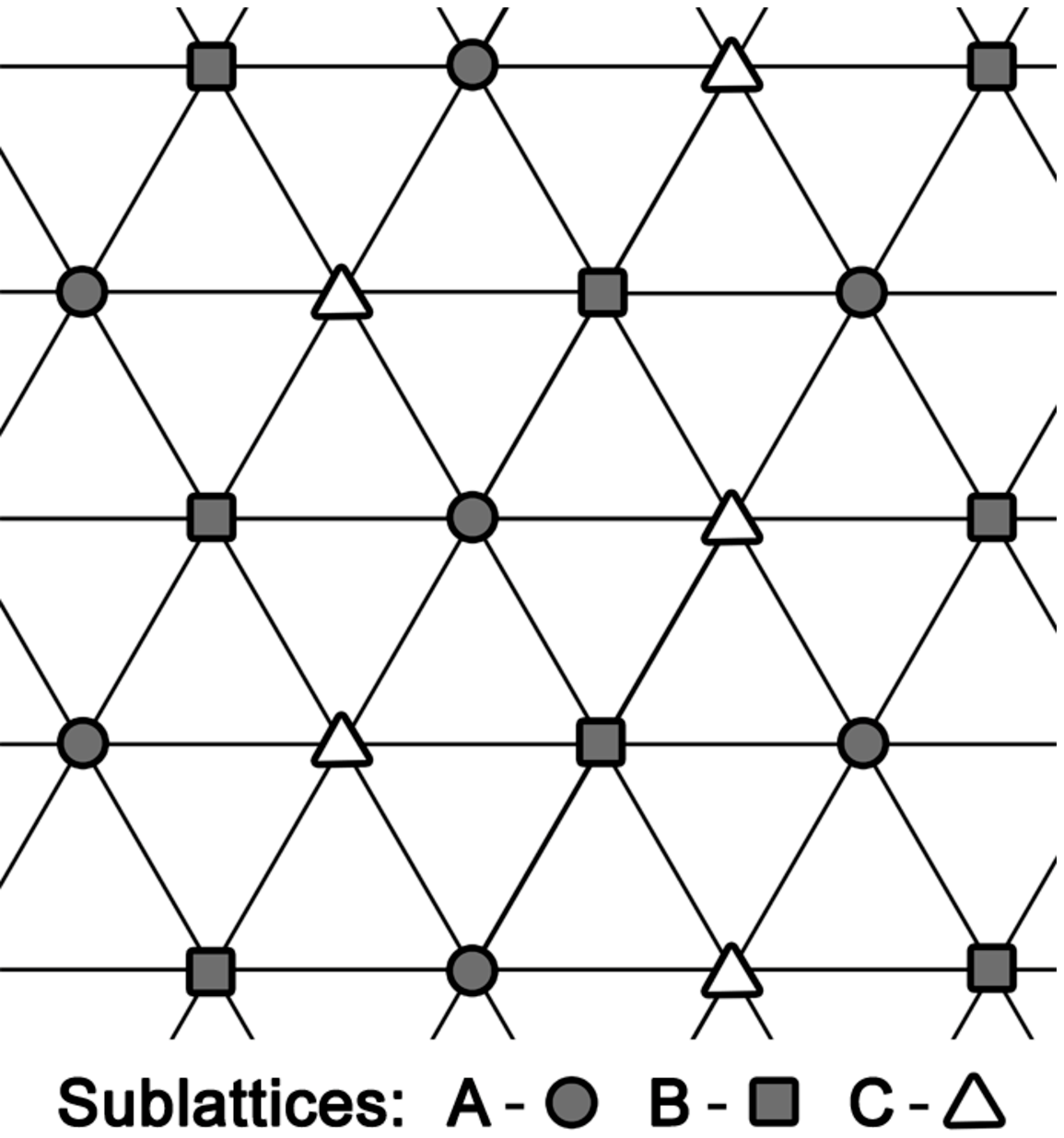}
\caption{Triangular lattice partition into three sublattices A,B and C. The shaded symbols signify the selective dilution case when two sublattices A and B are diluted, i.e., ${\bf p}=(p_{\mathrm{A}},p_{\mathrm{B}},1)$.}
\label{fig:lattice}
\end{figure}

Let us consider the Ising model on a triangular lattice, described by the Hamiltonian
\begin{equation}
\label{Hamiltonian}
H=-J\sum_{\langle i,j \rangle}\xi_{i}\xi_{j}S_{i}S_{j},
\end{equation}
where $S_{i}=\pm1$ are the Ising spin variables, $J<0$ is the exchange interaction coupling, and $\langle i,j \rangle$ is the sum extending over all nearest neighbor (NN) pairs. $\xi_{i}$ are quenched, uncorrelated random variables chosen to be equal to $1$ with probability $p$ when the site $i$ is occupied by a magnetic atom and $0$ with probability $1-p$ otherwise. Then $p$ represents the mean concentration of magnetic atoms. \\  
\hspace*{5mm} A uniformly diluted system, in which the vacancies are evenly distributed over the entire lattice, was investigated in the presence of an external field in our recent study~\cite{zuko} by the use of an effective field theory (EFT) with correlations (for review see, e.g.,~\cite{kane}). In order to include the geometrical frustration effects within EFT, the triangular lattice was decomposed into three interpenetrating sublattices A, B and C (see Fig.~\ref{fig:lattice}), such a way that spins on one sublattice can only interact with their NNs on the other two sublattices. Thus all the NN interactions are accounted for and the frustration results from the effort to simultaneously satisfy all the mutual antiferromagnetic intersublattice couplings. It should be noted here that the EFT with correlations is based on the differential operator technique introduced into the exact Ising spin identity and is almost as simply formulated as the standard mean-field theory. Another advantage of this method is that it enables systematic inclusion of the effects of spin correlations.  However, in this work we adopt the simplest approximation which neglects correlations between different spins but it takes the single-site kinematic relations exactly into account through the Var der Waerden identity. It is worthwhile noticing that this approach gives the same values of the critical temperature for the two- and three-dimensional Ising ferromagnets as the lowest linear approximation within the modified effective-field theory~\cite{kame} in the so called P scheme. Consequently, the present EFT yields, for example, a nonzero critical concentration for quenched diluted systems, the lack of order in the one-dimensional Ising ferromagnet, and the occurrence of order in the two-dimensional case with the critical temperature improved over the usual mean-field theory~\cite{kane} (for another advanced effective-field methods see also, e.g., Ref.~\cite{kato}). For the present frustrated pure system, in contrast to the mean-field theory, our simple EFT approach~\cite{zuko} correctly reproduced no long-range order behavior down to zero temperature in zero field and a fairly accurate phase diagram in finite fields\footnote{Comparison can be made with some other more accurate techniques, such as Monte Carlo and renormalization-group studies \cite{metcalf,schick,netz}}.\\
\hspace*{5mm} In the present paper, we generalize the study done in Ref.~\cite{zuko} by considering the concentration $p$ sublattice-dependent and hence characterized by a three-component vector ${\bf p}=(p_{\mathrm{A}},p_{\mathrm{B}},p_{\mathrm{C}})$. Following the procedure described in Ref.~\cite{zuko}, the A-sublattice magnetization per site in zero field can be calculated by taking a configurational average of the expression
\begin{align}
\label{conf_sub}
\xi_{i}^{\mathrm{A}}\langle S_{i}^{\mathrm{A}}\rangle=&\xi_{i}^{\mathrm{A}}\Big\langle
\prod_{j=1}^{z_{\mathrm{AB}}}[\xi_{j}^{\mathrm{B}}\cosh(\beta JD)+\xi_{j}^{\mathrm{B}}S_{j}^{\mathrm{B}}\sinh(\beta JD)+1-\xi_{j}^{\mathrm{B}}]\\ \nonumber
&\times\prod_{k=1}^{z_{\mathrm{AC}}}[\xi_{k}^{\mathrm{C}}\cosh(\beta JD)+\xi_{k}^{\mathrm{C}}S_{k}^{\mathrm{C}}\sinh(\beta JD)+1-\xi_{k}^{\mathrm{C}}]\Big\rangle\tanh(x)|_{x=0},
\end{align}
where $z_{\mathrm{AB}},z_{\mathrm{AC}}$ are the numbers of NNs of the spin $S_{i}^{\mathrm{A}}$ from the sublattice A that belong to the sublattices B and C, respectively, $\beta=1/k_{B}T$ and $D=\partial/\partial x$ is the differential operator. Similar expressions for the sublattices B and C can be obtained from Eq.~(\ref{conf_sub}) by cyclic permutation of the indices A, B and C and for each sublattice considering an appropriate number of NNs on the remaining sublattices (in our case $z_{\mathrm{XY}}=3$, for all pairs X, Y = A, B or C). Then, by the use of the decoupling approximation for the thermal averaging~\cite{zuko} and performing the configurational averaging of these equations for the respective sublattices leads to a system of coupled equations for the averaged sublattice magnetizations per site in the form
\begin{equation}
\begin{array}{l}
		\label{sub_mag_sel}
		m_{\mathrm{A}} = p_{\mathrm{A}} \left( a_{\mathrm{B}} + b m_{\mathrm{B}} \right)^3 \left( a_{\mathrm{C}} + b m_{\mathrm{C}} \right)^3 \tanh \left( x \right) |_{x=0}, \\
				m_{\mathrm{B}} = p_{\mathrm{B}} \left( a_{\mathrm{C}} + b m_{\mathrm{C}} \right)^3 \left( a_{\mathrm{A}} + b m_{\mathrm{A}} \right)^3 \tanh \left( x \right) |_{x=0}, \\
				m_{\mathrm{C}} = p_{\mathrm{C}} \left( a_{\mathrm{A}} + b m_{\mathrm{A}} \right)^3 \left( a_{\mathrm{B}} + b m_{\mathrm{B}} \right)^3 \tanh \left( x \right) |_{x=0},
\end{array}
\end{equation}
where $a_{\mathrm{X}} = 1 - p_{\mathrm{X}} + p_{\mathrm{X}} \cosh \left( \beta J D \right)$, with X = A, B or C, and $b = \sinh \left( \beta J D \right)$. Using the differential operator relation $\exp(\alpha D)f(x)=f(x+\alpha)$, Eqs.~(\ref{sub_mag_sel}) can be expressed explicitly. \\
\hspace*{5mm} In zero field, the critical temperature can be determined as follows. Using the fact that in the vicinity of the second-order phase transition the sublattice magnetizations $m_{\mathrm{X}}$ (X = A, B or C) are very small, Eqs.~(\ref{sub_mag_sel}) can be expanded and linearized, which leads to a system of homogeneous linear equations
\begin{equation}
{\bf U m = 0}, 
\label{sle}
\end{equation}
where
\begin{equation}
							{\bf U} = \left( 
\begin{array}{ccc}
-1 & 3 p_{\mathrm{A}} a_{\mathrm{B}}^2 a_{\mathrm{C}}^3 b \tanh \left( x \right) |_{x=0} & 3 p_{\mathrm{A}} a_{\mathrm{B}}^3 a_{\mathrm{C}}^2 b \tanh \left( x \right) |_{x=0} \\
3 p_{\mathrm{B}} a_{\mathrm{A}}^2 a_{\mathrm{C}}^3 b \tanh \left( x \right) |_{x=0} & -1 & 3 p_{\mathrm{B}} a_{\mathrm{A}}^3 a_{\mathrm{C}}^2 b \tanh \left( x \right) |_{x=0} \\
3 p_{\mathrm{C}} a_{\mathrm{A}}^2 a_{\mathrm{B}}^3 b \tanh \left( x \right) |_{x=0} & 3 p_{\mathrm{C}} a_{\mathrm{A}}^3 a_{\mathrm{B}}^2 b \tanh \left( x \right) |_{x=0} & -1 \\
\end{array} 	
							\right)
		\label{matrixKoef}
\end{equation}
and ${\bf m} = \left( m_{\mathrm{A}}, m_{\mathrm{B}}, m_{\mathrm{C}} \right)^{\rm T}$. The critical (or N\'{e}el) temperature $T_N$ can be established as a function of the sublattice concentration vector ${\bf p}=(p_{\mathrm{A}},p_{\mathrm{B}},p_{\mathrm{C}})$ by finding a nontrivial solution of Eqs. (\ref{sle}), i.e., by solving the secular equation 
\begin{equation}
	\det {\bf U} = 0.
\label{phaseDiag}
\end{equation}  

\section{Results and discussion}
In the following, we will consider a selectively diluted case such that two (let us say A and B) sublattices are diluted while the remaining sublattice (C) is occupied with purely magnetic ions, i.e., the case of ${\bf p}=(p_{\mathrm{A}},p_{\mathrm{B}},1)$, where $0 \le p_{\mathrm{A}}, p_{\mathrm{B}} \le 1$. The special case of ${\bf p}=(p_{\mathrm{A}},1,1)$, when only one sublattice is diluted, was investigated by hard-spin mean-field theory (HSMFT)~\cite{kaya} and Monte Carlo (MC) simulations~\cite{robi}. Compared to the regular mean-field theory, HSMFT conserves frustration since, instead of a uniform magnetization, it considers local magnetizations at each of N sites which feel the antialigning field due to the full (i.e., hard) spin of each of its neighbors. A further approximation (FA) of HSMFT still incorporates frustration, however, the approach is simplified by imposing sublatticewise uniformity. Thus, FA HSMFT leads to a set of only three instead of N coupled equations. These HSMFT and MC simulation works have shown that below some threshold concentration $p_{A}^{c}$, the nondiluted sublattices develop a long-range order (LRO) with nonvanishing magnetizations $m_{\mathrm{B}}=-m_{\mathrm{C}}$ that do not saturate as the temperature approaches zero. On the other hand, the magnetization of the diluted sublattice $m_{\mathrm{A}}$ remains zero at any concentration.
\begin{figure}[t!]
\centering
   \includegraphics[scale=0.7]{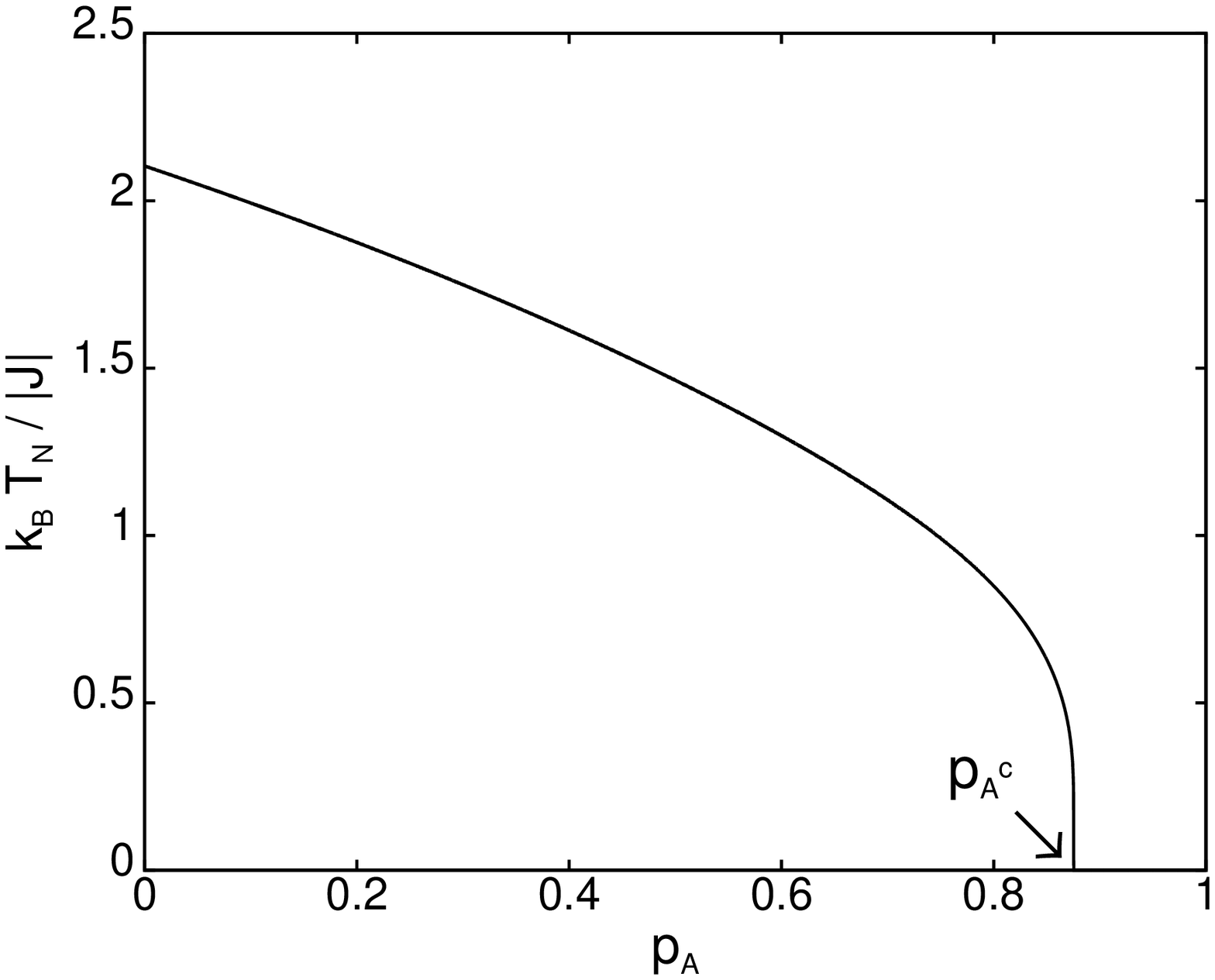}
\caption{Transition temperature as a function of the concentration $p_{\mathrm{A}}$, for ${\bf p}=(p_{\mathrm{A}},1,1)$.}
\label{fig:phase_pA_h0}
\end{figure}
\begin{figure}[t!]
\centering
   \includegraphics[scale=0.7]{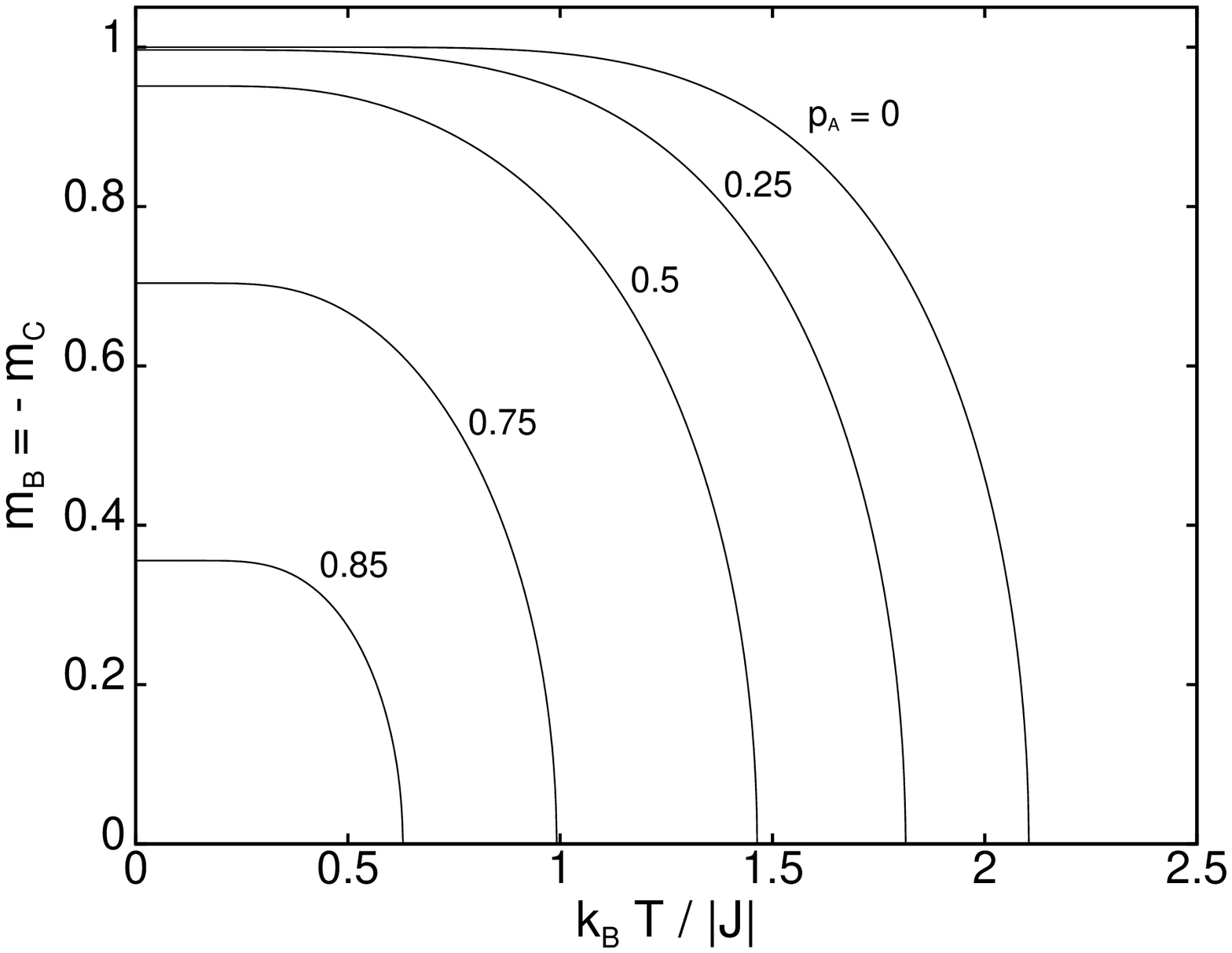}
\caption{Temperature variation of the nondiluted sublattice magnetizations $m_{\mathrm{B}}=-m_{\mathrm{C}}$, for different values of $p_{\mathrm{A}}$, for the case ${\bf p}=(p_{\mathrm{A}},1,1)$.}
\label{fig:zero_field_m}
\end{figure}
A similar picture is also observed in our EFT approach. The resulting phase diagram is plotted in Fig.~\ref{fig:phase_pA_h0}, in which the area below the curve represents the LRO phase. This diagram can be compared with similar plots obtained from HSMFT and FA HSMFT, shown in Fig. 3 of Ref.~\cite{kaya}. In both EFT and HSMFT approaches the transition temperature decreases with increasing concentration, albeit, the EFT results seem to be quantitatively closer to those obtained from FA HSMFT rather than HSMFT. Namely, the EFT critical value of $p_{\mathrm{A}}^c = 0.8753$, above which no LRO exists in either of the sublattices, matches well with the one obtained from FA HSMFT ($p_{\mathrm{A}}^c = 0.875$) but it is lower than those obtained from HSMFT ($p_{\mathrm{A}}^c = 0.958$) and MC simulations ($p_{\mathrm{A}}^c \approx 0.95$)~\cite{robi}, respectively. In the limit of $p_{\mathrm{A}}=0$ the EFT critical temperature of the honeycomb lattice model of $k_BT_N/|J|=2.1037$ is accurately recovered (see, e.g., Ref.~\cite{tagg} where also comparison with some other methods is provided). In Fig.~\ref{fig:zero_field_m} we plot a temperature variation of the nondiluted sublattice magnetizations $m_{\mathrm{B}}=-m_{\mathrm{C}}$ for different values of $p_{\mathrm{A}}$. By comparing it with Fig. 1(a) of Ref.~\cite{kaya}, one can see a qualitatively similar behavior: the sublattice magnetizations do not saturate at zero-temperature, except for the full dilution (honeycomb lattice) limit, and not only the transition temperatures (as already seen in Fig.~\ref{fig:phase_pA_h0} shown above) but also the values of the sublattice magnetizations at zero temperature decrease with increasing concentration $p_{\mathrm{A}}$. \\
\hspace*{5mm} Now let us see how this picture will change if the dilution is extended to two sublattices, i.e., if ${\bf p}=(p_{\mathrm{A}},p_{\mathrm{B}},1)$, where $0 \le p_{\mathrm{A}}, p_{\mathrm{B}} \le 1$. The corresponding phase diagram in the $(p_{\mathrm{A}}, p_{\mathrm{B}})$ plane is presented in Fig.~\ref{fig:phase_pA_pB_h0}. We note that with respect to the sublattice magnetizations, the critical surface shown in Fig.~\ref{fig:phase_pA_pB_h0} represents the temperatures above which all the sublattice magnetizations vanish. However, in case of ${\bf p}=(p_{\mathrm{A}},1,1)$ presented above, the system can display a partially ordered phase with one (diluted) sublattice magnetization equal to zero even below the critical surface. The above mentioned special case of ${\bf p}=(p_{\mathrm{A}},1,1)$, shown in Fig.~\ref{fig:phase_pA_h0}, is represented by the curve of $p_{\mathrm{B}}=1$ and varying $p_{\mathrm{A}}$. When we also start diluting the sublattice B we observe overall decrease of the transition temperature and until $p_{\mathrm{B}}=p_{\mathrm{A}}$ also decrease of the value of $p_{\mathrm{A}}^c$ from $0.8753$ at $p_{\mathrm{B}}=1$ down to $0.7640$ at $p_{\mathrm{B}}=p_{\mathrm{A}}$. Moreover, we have found that for $p_{\mathrm{B}} \in (0.7640,0.8753)$ the system shows the reentrant behavior. Namely, at a fixed value of $p_{\mathrm{B}}$ by varying the value of $p_{\mathrm{A}}$ the system passes from the ordered phase to the disordered one and then again to the ordered one. This behavior can be more clearly seen in Fig.~\ref{fig:phase_pA_pB_h0_GS} in which the ground-state phase diagram is depicted. The latter can be obtain from Eq.~(\ref{phaseDiag}) in zero-temperature limit. For $p_{\mathrm{B}} \in (0.4280,0.7640)$, LRO persists for any value of $p_{\mathrm{A}}$, however, at $p_{\mathrm{B}}=p_{\mathrm{A}}$ the transition temperature curve changes from decreasing to increasing. For $p_{\mathrm{B}} \in (0.2976,0.4280)$ the system again displays the reentrant behavior with LRO disappearing within a certain range of the concentration $p_{\mathrm{A}}$. Upon further decrease of $p_{\mathrm{B}}$ there is no LRO at lower values of $p_{\mathrm{A}}$ but it appears at some threshold value $p_{\mathrm{A}}^c$, which decreases down to $p_{\mathrm{A}}^c=0.2976$ at $p_{\mathrm{B}}=0$. The limiting case of $p_{\mathrm{B}}=0$ (or equivalently $p_{\mathrm{A}}=0$) corresponds to the honeycomb lattice. Therefore, if we consider the case of $p_{\mathrm{A}}=0$, the dependence of the critical temperature on $p_{\mathrm{B}}$ represents the phase diagram of the selectively diluted Ising antiferromagnet on the honeycomb lattice and the value $p_{\mathrm{B}}^c=0.2976$ is the percolation threshold, i.e., the critical concentration below which LRO cannot exist due to the absence of a spanning cluster of magnetic ions (Fig.~\ref{fig:phase_pB_pA0}). We note that the phase diagram is symmetric along the diagonal and therefore the above arguments also apply when $p_{\mathrm{A}}$ and $p_{\mathrm{B}}$ are interchanged.\\
\hspace*{5mm} At this point, it is interesting to examine the frustration effect on the percolation behavior by comparing it with its nonfrustrated counterpart, i.e., with the ferromagnetic case. The selective site percolation problem in an Ising ferromagnet on triangular and square lattices with several kinds of sublattice structures has been extensively studied by Idogaki and Ury\^{u}~\cite{ido1,ido2,ido3}, employing a renormalization group technique. However, it is difficult to assess the frustration effect by a direct comparison of the percolation thresholds from our study with those obtained in Refs.~\cite{ido1,ido2,ido3} due to generally different sublattice structures considered as well as different techniques used. Therefore, we made additional calculations for the current selectively diluted model within EFT but considering the exchange interaction $J>0$, i.e., for the triangular Ising ferromagnet, and the results are presented by the dashed curve in Fig.~\ref{fig:phase_pA_pB_h0_GS}. Naturally, for the ferromagnetic case there is no disordered region at high concentrations and the low concentration disordered region is much smaller than for the frustrated antiferromagnetic case. As can be expected, the biggest difference between the two cases is along the diagonal ($p_{\mathrm{A}}=p_{\mathrm{B}}$), where the frustration in the former case is the highest, and vanishes in the limits of $p_{\mathrm{A}}=0$ and $p_{\mathrm{B}}=0$, where there is no frustration. 
\begin{figure}[t!]
\centering
   \includegraphics[scale=0.8]{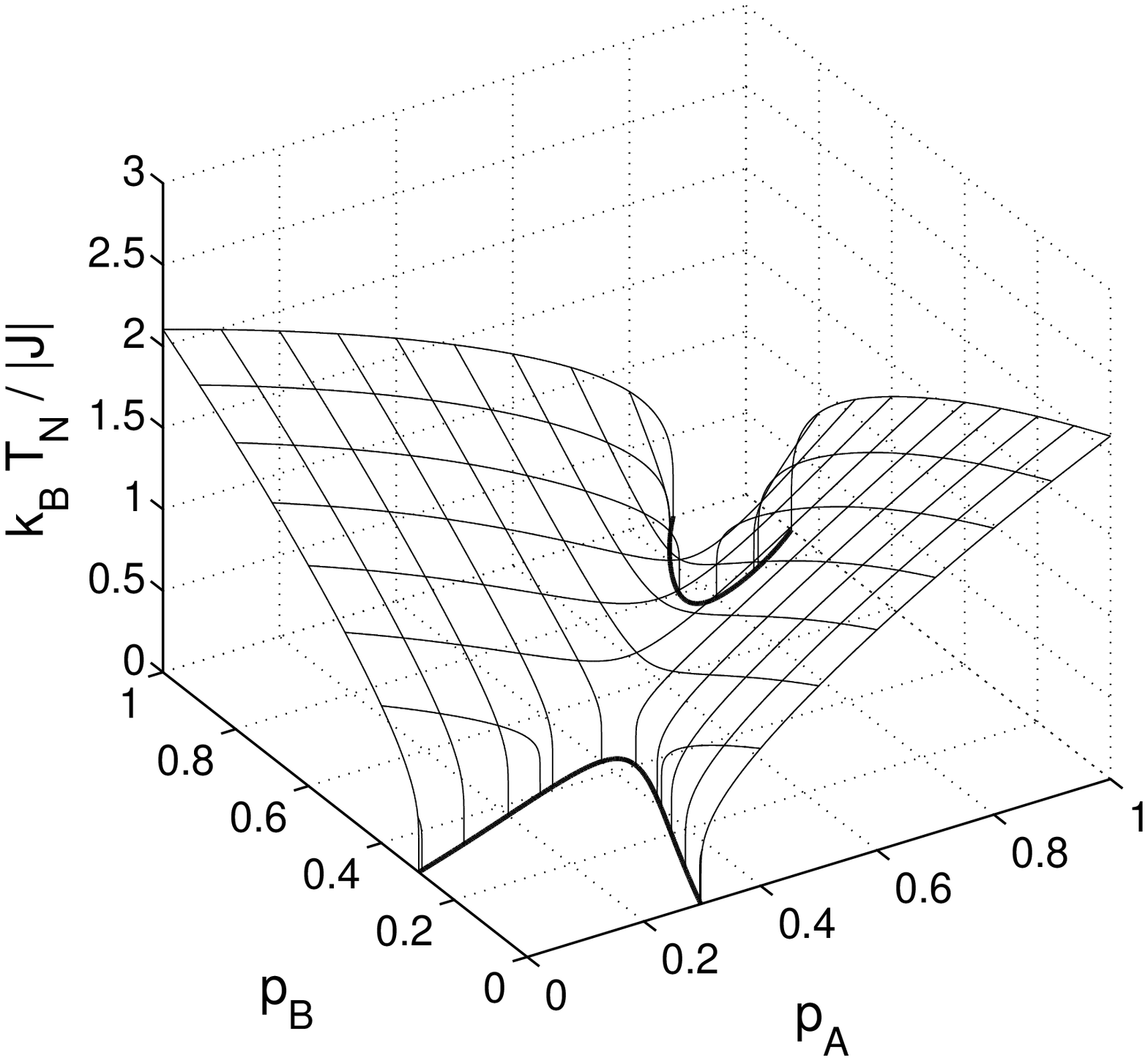}
\caption{Transition temperature as a function of the concentrations $p_{\mathrm{A}}$ and $p_{\mathrm{B}}$, for ${\bf p}=(p_{\mathrm{A}},p_{\mathrm{B}},1)$. The bold lines outline the zero-temperature limit.}
\label{fig:phase_pA_pB_h0}
\end{figure}
\begin{figure}[t!]
\centering
   \includegraphics[scale=0.8]{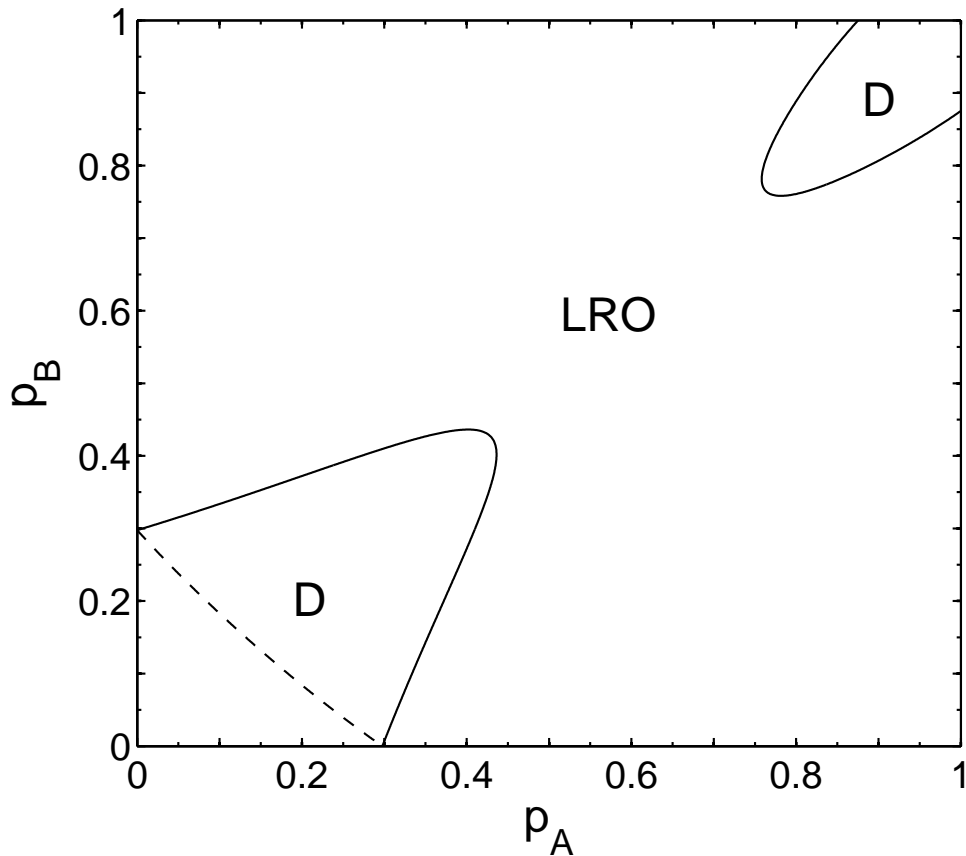}
\caption{Ground-state phase diagram in $p_{\mathrm{A}}-p_{\mathrm{B}}$ concentration plane, for ${\bf p}=(p_{\mathrm{A}},p_{\mathrm{B}},1)$. LRO and D denote the long-range ordered and disordered phases, respectively. The dashed curve corresponds to the nonfrustrated ferromagnetic case.}
\label{fig:phase_pA_pB_h0_GS}
\end{figure}
\begin{figure}[t!]
\centering
   \includegraphics[scale=0.7]{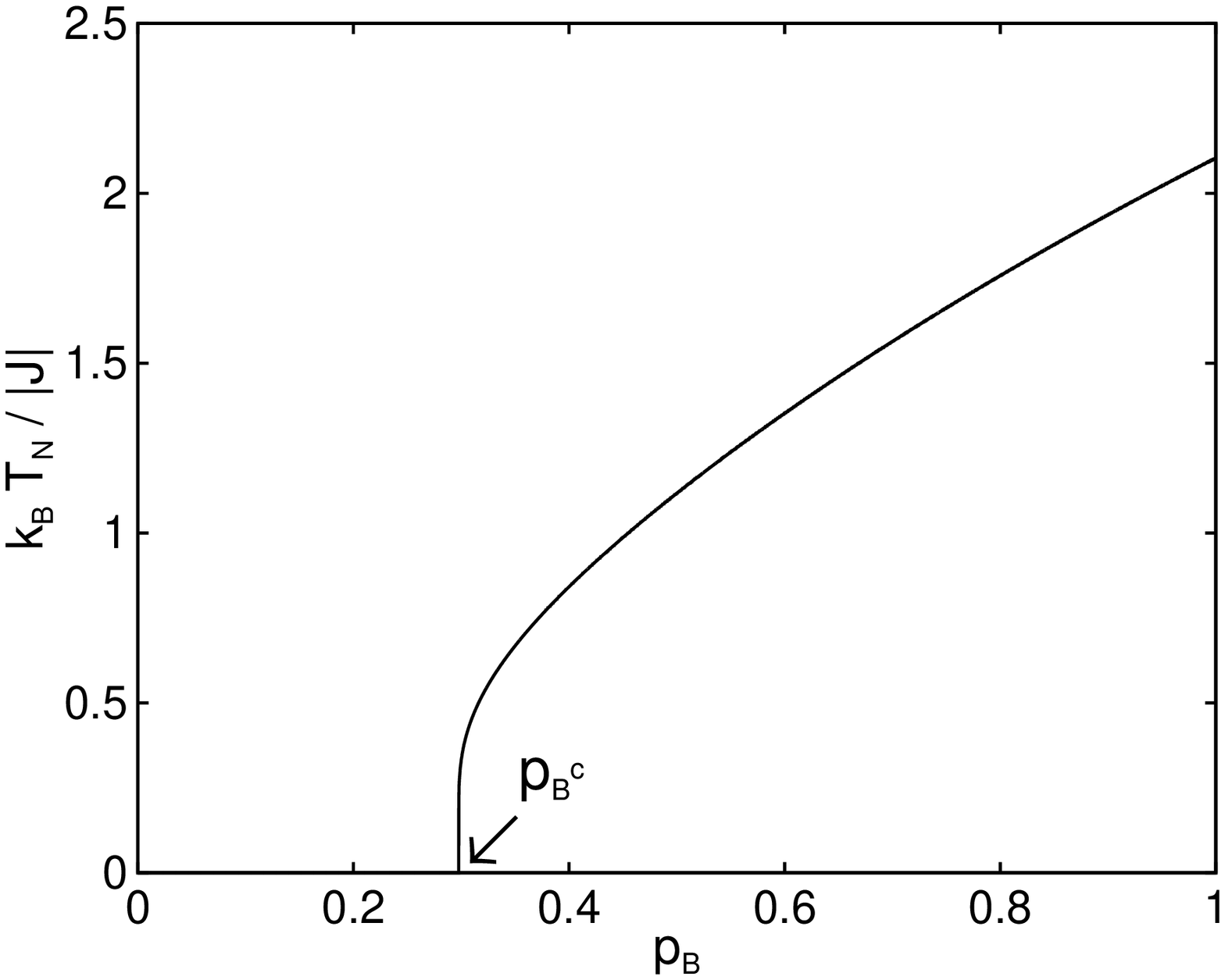}
\caption{Transition temperature as a function of the concentration $p_{\mathrm{B}}$, for ${\bf p}=(0,p_{\mathrm{B}},1)$. This dependence represents the phase diagram of the selectively diluted antiferromagnet on a honeycomb lattice.}
\label{fig:phase_pB_pA0}
\end{figure}
\hspace*{5mm} For the case of ${\bf p}=(p_{\mathrm{A}},1,1)$, investigated in Refs.~\cite{kaya,robi} as well as in the current study, it has been shown that below the threshold value of $p_{\mathrm{A}}^c$ the nondiluted sublattice magnetizations become finite and the diluted sublattice magnetization always remains zero. Moreover, at zero temperature the nondiluted sublattice magnetizations do not saturate for any nonzero concentration. In order to shed more light on these phenomena, let us study the ground-state behavior of the current model with two sublattices diluted, i.e., the case of ${\bf p}=(p_{\mathrm{A}},p_{\mathrm{B}},1)$. It is well known that in the pure model the elementary triangular plaquette in the ground state is six-fold degenerate. The selective site dilution lifts the degeneracy and the system can show long-range order. The possible zero-temperature sublattice arrangements and their respective energies are presented in Table~\ref{tab:GSconf}. It is easy to verify that if both sublattices are diluted, i.e., $p_{\mathrm{A}}, p_{\mathrm{B}} < 1$, then the lowest energy is $e_3=p_{\mathrm{A}} p_{\mathrm{B}} - p_{\mathrm{A}} - p_{\mathrm{B}}$, and the corresponding configurations are those with spins on the diluted sublattices A and B oriented in one direction and spins on the nondiluted sublattice C in the opposite direction. On the other hand, if only one sublattice is diluted, e.g., $p_{\mathrm{A}} < 1$ and $p_{\mathrm{B}} = 1$,  then the lowest energy is $e_1=e_3=-1$, and the corresponding configurations are those with spins on the nondiluted sublattices B and C oriented antiparallel but spins on the diluted sublattice A can point in either direction with equal probability. This four-fold degeneracy explains why the diluted sublattice magnetization remains zero.\\
\begin{table}[t!]
\caption{Zero-temperature degeneracy with possible sublattice arrangements $\left(m_{\mathrm{A}},m_{\mathrm{B}},m_{\mathrm{C}}\right)$ and corresponding energies per site $\langle H \rangle/|J|N$ for ${\bf p}=(p_{\mathrm{A}},p_{\mathrm{B}},1)$.}
\label{tab:GSconf}
\centering
\begin{tabular}{c|ccc}
$n$ & 1 & 2 & 3 \\ \hline
$\left(m_{\mathrm{A}},m_{\mathrm{B}},m_{\mathrm{C}}\right)$ & 
			$\left(+,-,+ \right); \left(-,+,- \right)$ & 
			$\left(-,+,+ \right); \left(+,-,- \right)$ & 
			$\left(-,-,+ \right); \left(+,+,- \right)$ \\ \hline
$e_n=\frac{\langle H \rangle}{|J|N}$& 
			$ - p_{\mathrm{A}} p_{\mathrm{B}} + p_{\mathrm{A}} - p_{\mathrm{B}} $ & 
			$ - p_{\mathrm{A}} p_{\mathrm{B}} - p_{\mathrm{A}} + p_{\mathrm{B}} $ & 
			$ p_{\mathrm{A}} p_{\mathrm{B}} - p_{\mathrm{A}} - p_{\mathrm{B}} $ \\
\end{tabular}
\end{table}
\hspace*{5mm} The unsaturated magnetizations of the nondiluted sublattices for the ${\bf p}=(p_{\mathrm{A}},1,1)$ case, shown in Fig.~\ref{fig:zero_field_m}, apparently result from the presence of the frustration. As the dilution (frustration) increases (decreases), the magnetizations gradually reach the saturation value of 1. The initial increase is fast and for $p_{\mathrm{A}}=0.5$ about $96\%$ of the saturation value is reached. Nevertheless, with decreasing $p_{\mathrm{A}}$ the curve flattens out and the full saturation is only achieved at $p_{\mathrm{A}}=0$. This behavior is illustrated in Fig.~\ref{fig:mag_pA_pB1}, in which we plot the sublattice magnetizations at zero temperature. The results are in good agreement with those presented in Fig. 2(a) of Ref.~\cite{kaya}. Further, due to considerable variation of the frustration degree in the $(p_{\mathrm{A}},p_{\mathrm{B}})$ parameter space, it will be interesting to see similar plots also for the ${\bf p}=(p_{\mathrm{A}},p_{\mathrm{B}},1)$ case at different values of $p_{\mathrm{A}}$ and $p_{\mathrm{B}}$. The reached saturation level is better seen if instead of the sublattice magnetizations {\it per site} $m_{\mathrm{X}}$ we plot the sublattice magnetizations {\it per spin} $\mu_{\mathrm{X}}=m_{\mathrm{X}}/p_{\mathrm{X}}$, where X = A,B or C. Note that in Fig.~\ref{fig:mag_pA_pB1} $\mu_{\mathrm{X}}=m_{\mathrm{X}}$. In Fig.~\ref{fig:zero_temp_mag} we present the variations of the zero-temperature sublattice magnetizations per spin in different regions of the $(p_{\mathrm{A}},p_{\mathrm{B}})$ plane. For $p_{\mathrm{B}}=0.8$, as expected from Figs.~\ref{fig:phase_pA_pB_h0} and \ref{fig:phase_pA_pB_h0_GS}, we observe the reentrant behavior featuring a certain region of the values of $p_{\mathrm{A}}$ in which $\mu_{\mathrm{X}}=0$, for X = A,B or C (Fig.~\ref{fig:zero_temp_mag_a}). Outside of this highly frustrated region, in accordance with the ground-state analysis presented in Table~\ref{tab:GSconf}, the diluted sublattices A and B have nonzero magnetizations with the signs opposite to that of the nondiluted sublattice C, and in the limit of $p_{\mathrm{A}}=1$ the magnetization of the remaining diluted sublattice B reduces to zero. Interestingly, in the other limit of $p_{\mathrm{A}} \to 0$, although the A sublattice magnetization per site $m_{\mathrm{A}}$ vanishes, the value of  $\mu_{\mathrm{A}}$ remains finite and drops discontinuously to zero only at $p_{\mathrm{A}} = 0$. Also, the magnetization of the nondiluted sublattice C falls short of saturation. A closer inspection of Fig.~\ref{fig:zero_temp_mag_a} reveals that the sublattice magnetization $\mu_{\mathrm{C}}$ does not reach its full saturation value of -1 even at $p_{\mathrm{A}} = 0$, i.e., for the selectively diluted honeycomb lattice system. Thus, the lack of saturation in this case cannot be ascribed to the frustration but rather to the presence of small amount of isolated C-ions or isolated clusters involving both C- and B-ions surrounded by vacancies which do not contribute to the magnetizations of the C- or both C- and B-sublattices, respectively. The above described phenomena can be observed even more pronouncedly for $p_{\mathrm{B}}=0.4$ (Fig.~\ref{fig:zero_temp_mag_b}). The more abundant presence of the ``loose'' clusters involving also B-ions with decreasing $p_{\mathrm{B}}$ is reflected in the decreasing value of $\mu_{\mathrm{B}}$ at $p_{\mathrm{A}} = 0$. Here we note that in both Fig.~\ref{fig:zero_temp_mag_a} and (b) the source of the unsaturation in the limit of $p_{\mathrm{A}} = 0$ is different from that observed in the limit of $p_{\mathrm{A}} = 1$. In the former case the system is nonfrustrated (selectively diluted honeycomb lattice) and the lack of saturation can be ascribed solely to the presence of the "loose" isolated C-ions or clusters of C- and B-ions, which do not align with the spanning clusters. On the other hand, in the latter case the system is frustrated (selectively diluted triangular lattice) and the unsaturation appears in the nondiluted sublattices. Since the random dilution of the A-sublattice relieves frustration locally, the alignment of local magnetizations in the nonfrustrated sublattices proceeds in a gradual manner. Therefore, one can expect the presence of some amount of antialigned local magnetizations that cause the lack of saturation until the A-sublattice is completely removed by dilution. For the intermediate concentrations $0<p_{\mathrm{A}}<1$ the unsaturation effect is amplified by the presence of both sources. Namely, by increasing (decreasing) $p_{\mathrm{A}}$ from zero (one) the increasing frustration (dilution) magnifies the degree of unsaturation until the long-range order is destroyed in all the sublattices. In Fig.~\ref{fig:zero_temp_mag_c} we show the case of $p_{\mathrm{B}}=0$, i.e., the selectively diluted honeycomb lattice case. In this case the unsaturated magnetizations are due to isolated C-ions and clusters of C- and A-ions. If the sublattices A and B are diluted with equal probabilities, then the frustration is the least relieved and LRO with considerably suppressed values of the zero-temperature sublattice magnetizations appears only in a relatively small range of $0.4280 \le p_{\mathrm{A}}=p_{\mathrm{B}} \le 0.7640$ (Fig.~\ref{fig:zero_temp_mag_d}). On the other hand, the frustration is the most avoided if the dilution is performed such a way that $p_{\mathrm{A}}+p_{\mathrm{B}}=1$ (Fig.~\ref{fig:zero_temp_mag_e}). In this case LRO survives at any dilution. 

\begin{figure}[t!]
\centering
   \includegraphics[scale=0.7]{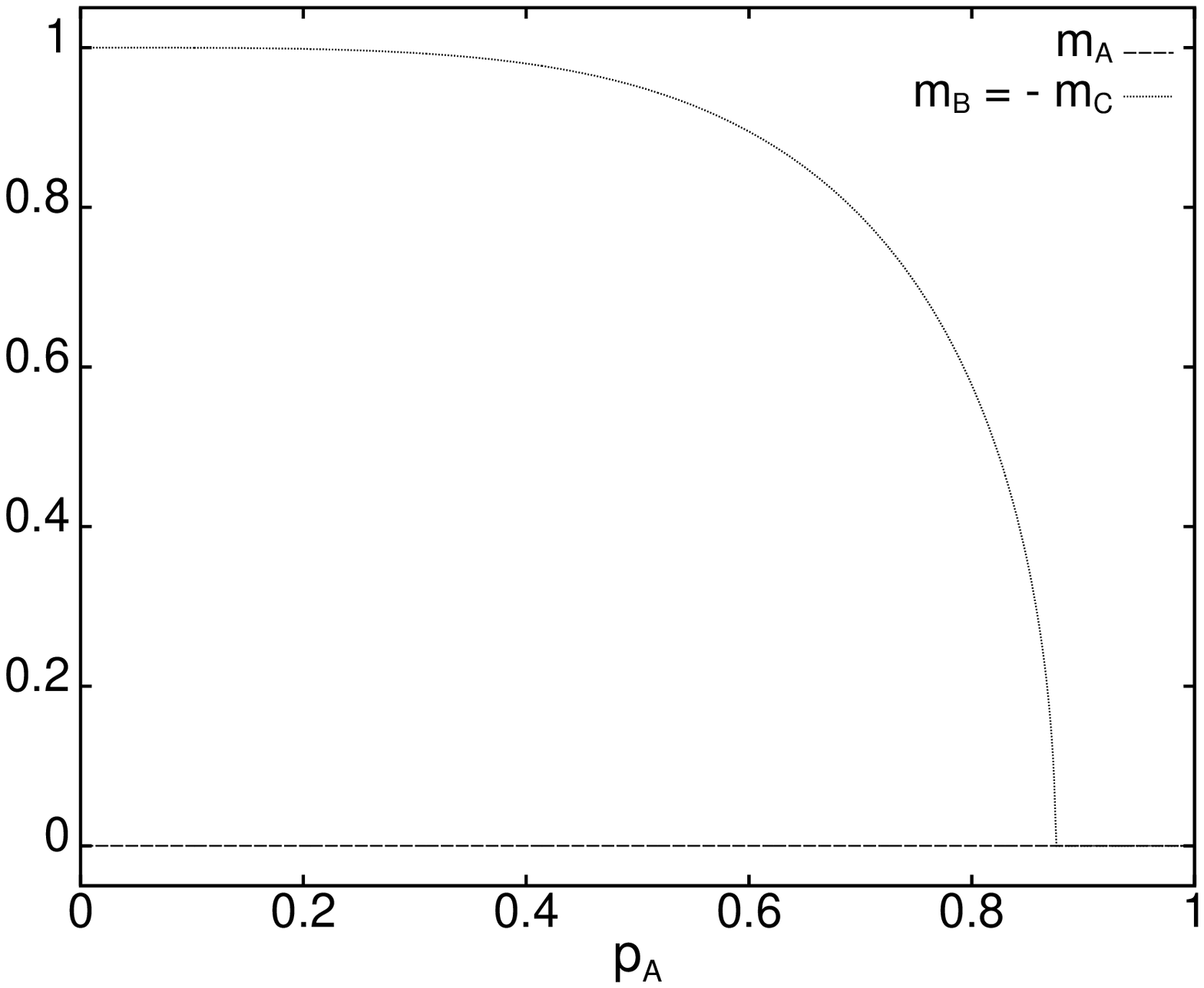}
\caption{Zero-temperature sublattice magnetizations for ${\bf p}=(p_{\mathrm{A}},1,1)$.}
\label{fig:mag_pA_pB1}
\end{figure}

\begin{figure}[ht]
  \centering\hspace*{-10mm}
    \subfigure[]{\label{fig:zero_temp_mag_a}\includegraphics[scale=0.45]{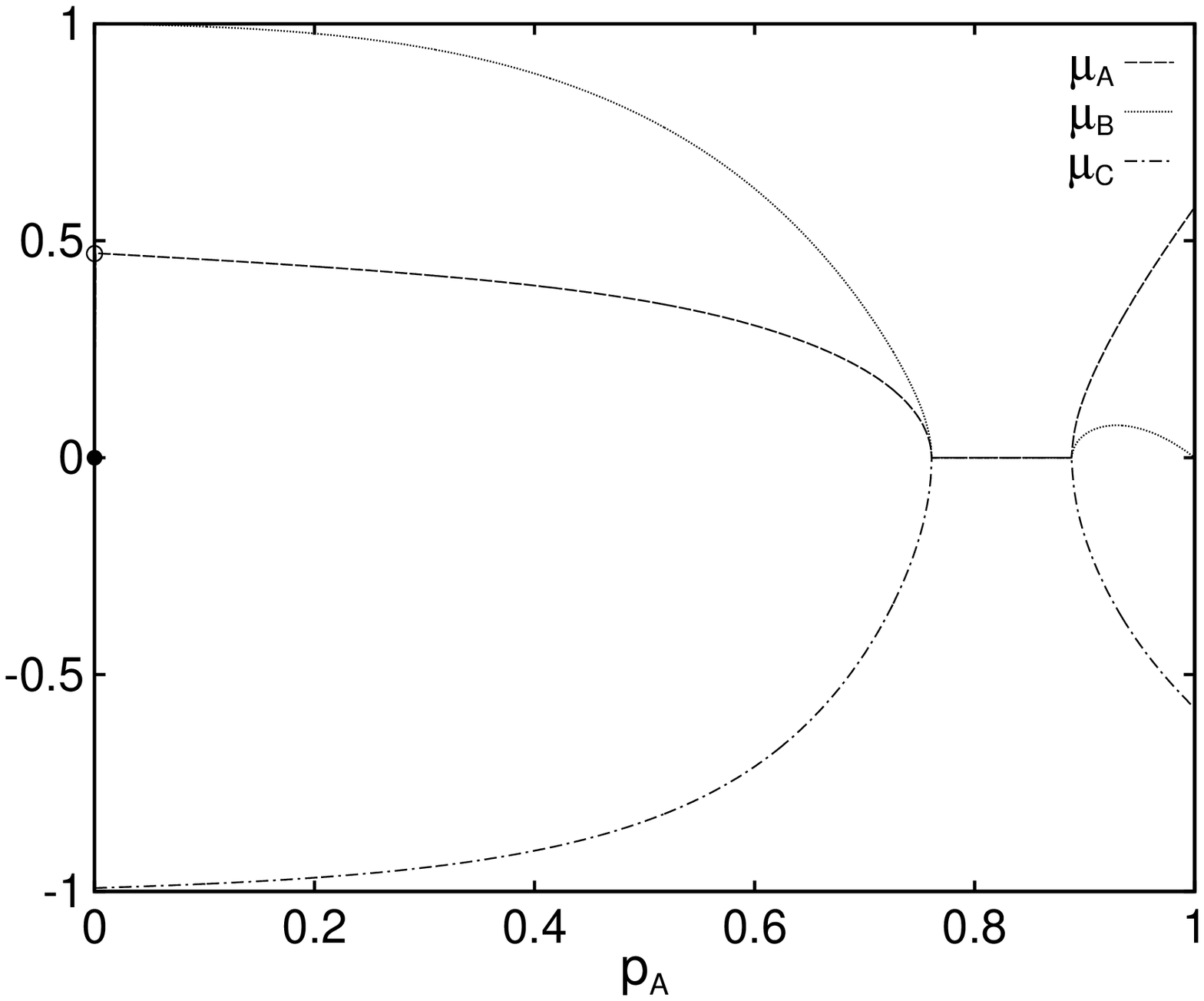}}\hspace*{-10mm}
    \subfigure[]{\label{fig:zero_temp_mag_b}\includegraphics[scale=0.45]{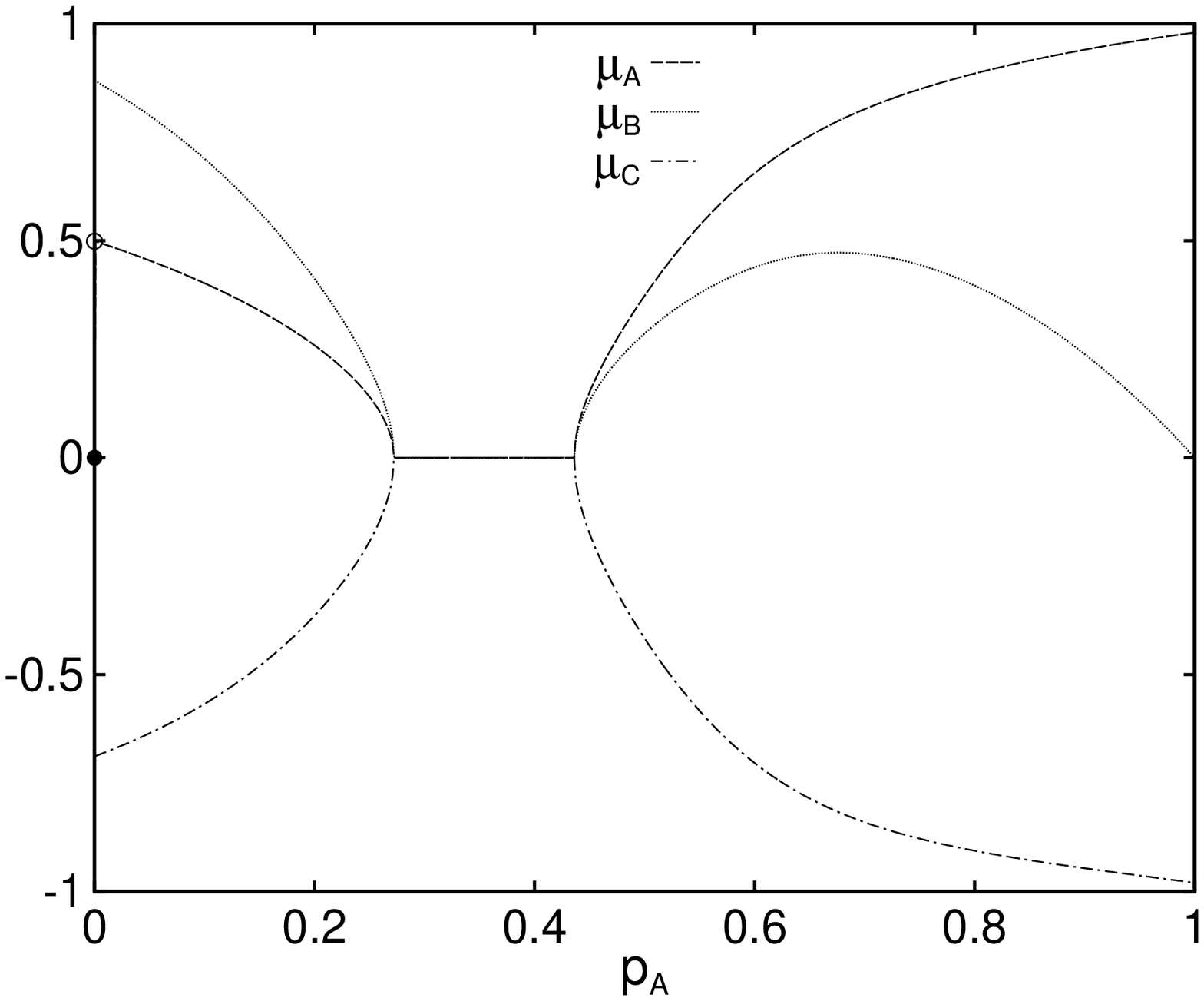}}\\ \vspace*{-3mm} \hspace*{-10mm}
    \subfigure[]{\label{fig:zero_temp_mag_c}\includegraphics[scale=0.45]{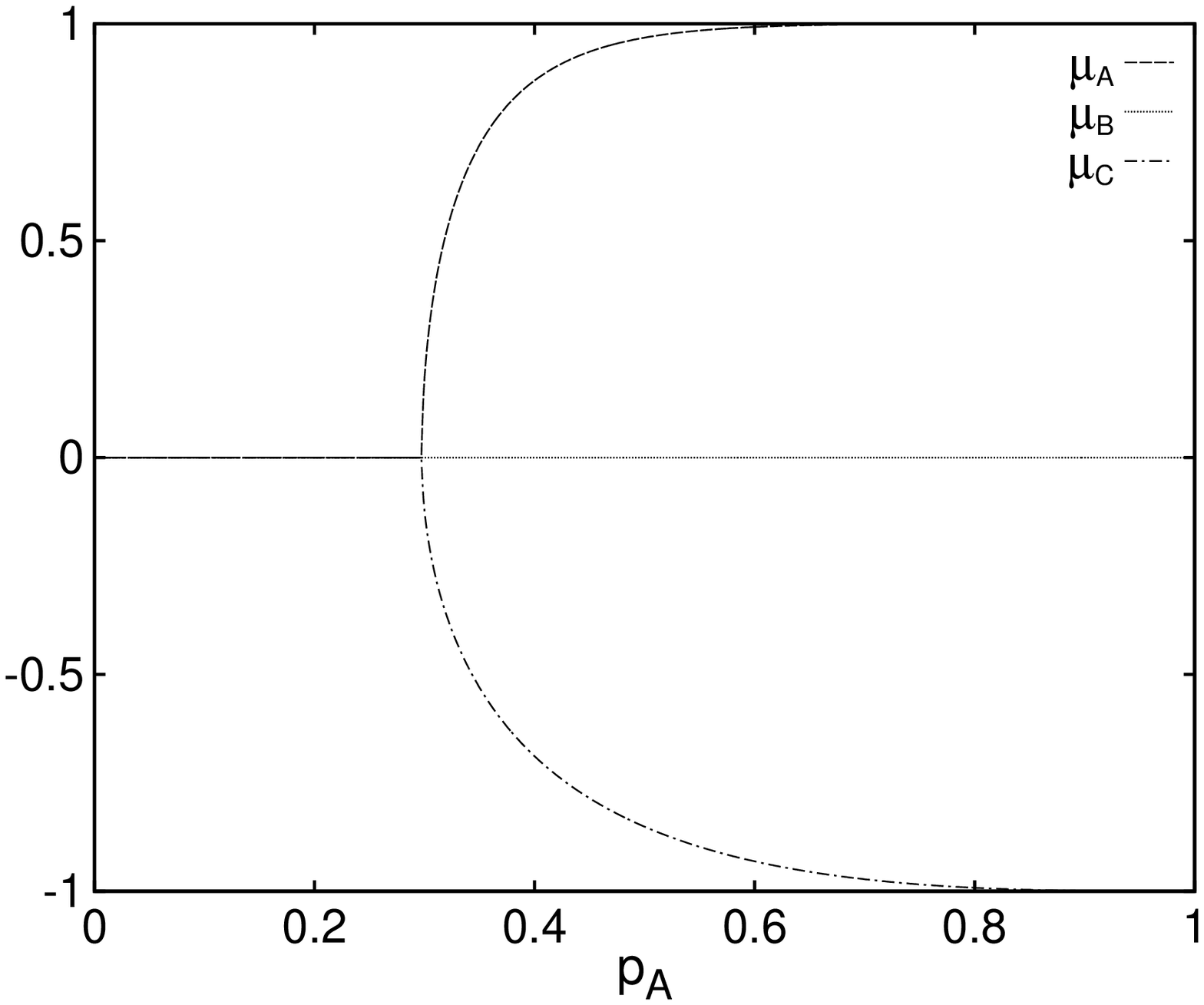}}\hspace*{-10mm}
    \subfigure[]{\label{fig:zero_temp_mag_d}\includegraphics[scale=0.45]{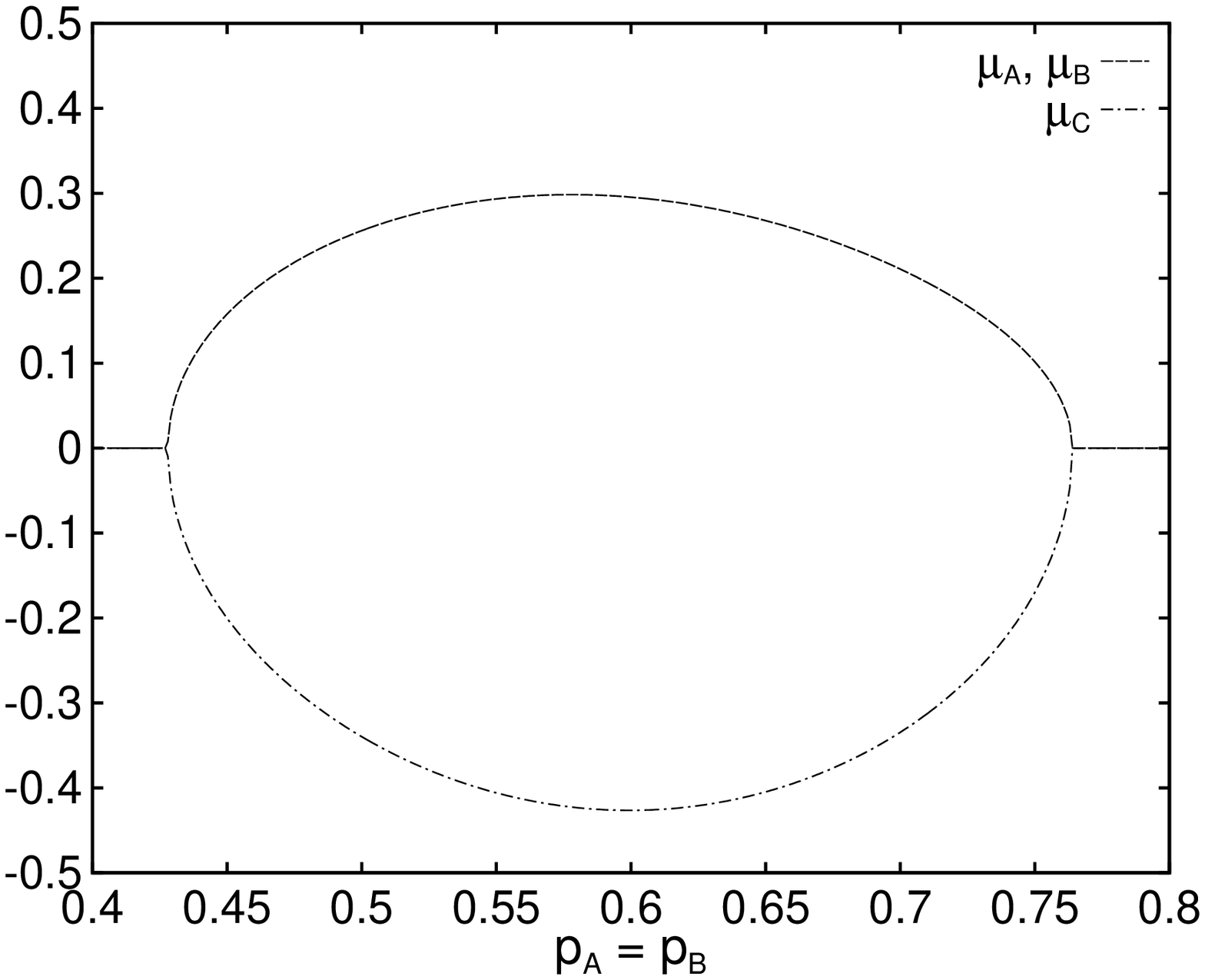}}\\ \vspace*{-3mm} \hspace*{-10mm}
    \subfigure[]{\label{fig:zero_temp_mag_e}\includegraphics[scale=0.45]{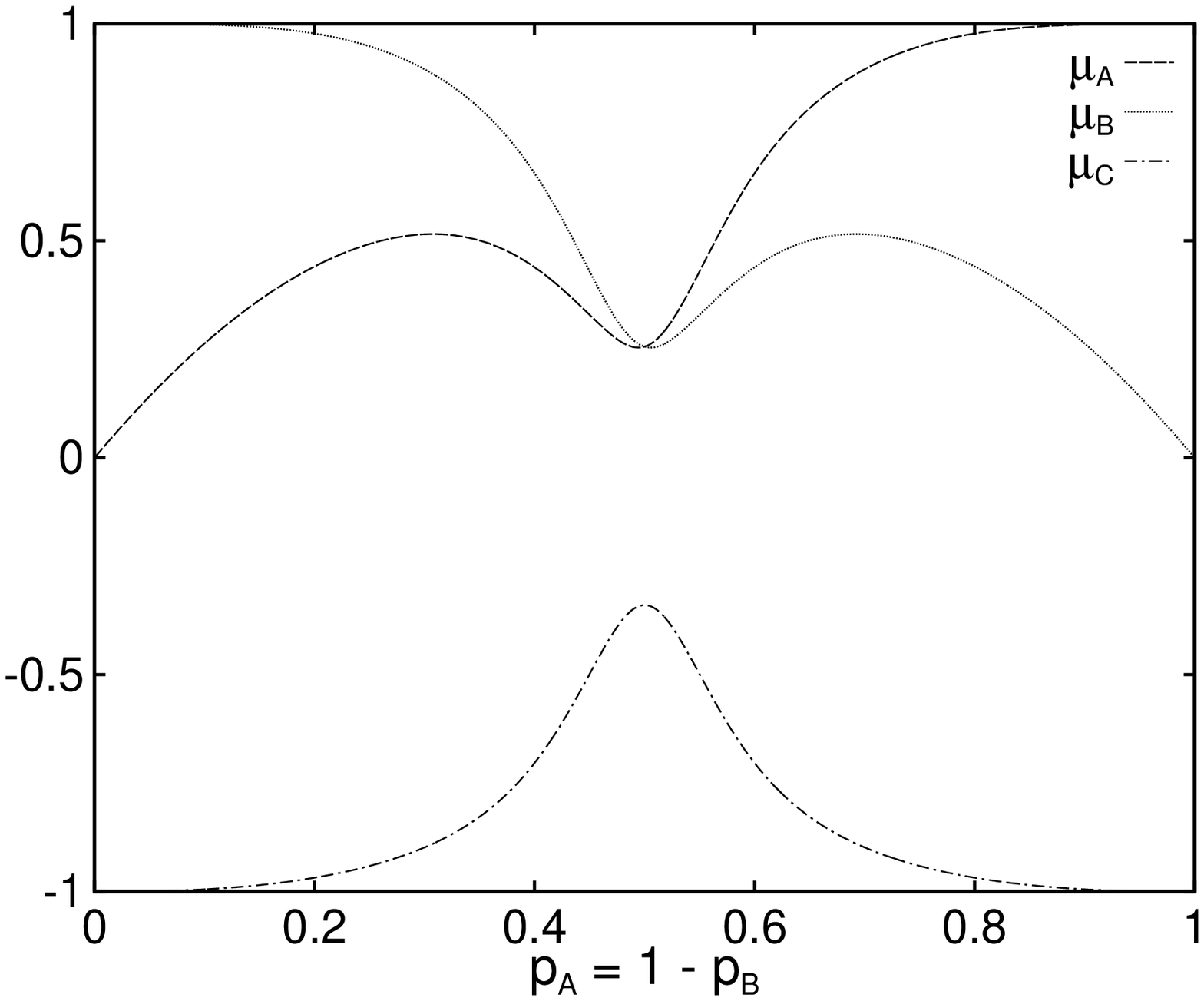}}\hspace*{-10mm}
  \caption{Zero-temperature sublattice magnetizations per spin for (a) $p_{\mathrm{B}}=0.8$, (b) $p_{\mathrm{B}}=0.4$, (c) $p_{\mathrm{B}}=0$, (d) $p_{\mathrm{A}}=p_{\mathrm{B}}$, and (e) $p_{\mathrm{A}}=1-p_{\mathrm{B}}$.}
  \label{fig:zero_temp_mag}
\end{figure}

\section{Summary and conclusions}
We studied the critical and ground-state behavior of a selectively site-diluted Ising antiferromagnet on triangular and honeycomb lattices within the framework of an effective-field theory with correlations. The selective dilution was carried out by random removal of magnetic ions from two of the three sublattices with different probabilities. In a special case when only one sublattice is diluted, our results showed a fairly good agreement with the previous studies by hard-spin mean-field theory~\cite{kaya} and Monte Carlo simulations~\cite{robi}. The extension to the system with two sublattices jointly diluted with different probabilities is an entirely new contribution. We established a three-dimensional phase diagram in the space of the sublattice concentrations of the triangular lattice. We identified one smaller disordered region at high sublattice concentrations in which a long-range order cannot occur due to high frustration. At low sublattice concentrations there is another disordered region in which a long-range order is absent due to the absence of a spanning cluster of magnetic ions. The two disordered regions are stretched along the diagonal owing to the increased frustration when the sublattice concentrations are about the same. These relatively small regions are surrounded by one larger region in which the sublattice concentrations are large enough to form a spanning cluster and the frustration is relieved enough to enable long-range ordering. We also studied the ground-state ordering and the effect of unsaturated sublattice magnetizations. The degree of unsaturation was found to be proportional to both the degree of frustration and dilution. However, due to compensating effects of the two factors (the increasing dilution decreases frustration), the highest unsaturation was observed at the intermediate dilution range, in the vicinity of the phase boundaries. The selectively diluted Ising antiferromagnet on the honeycomb lattice was obtained as a special case when one sublattice of the triangular lattice is completely diluted. We established the phase diagram including the percolation threshold below which no long-range order can exist. \\
\hspace*{5mm} In the present study we demonstrated how the selective dilution of one or two sublattices in the triangular lattice antiferromagnet can relieve frustration and cause long-range ordering, featuring some peculiar phenomena, such as reentrance and unsaturated sublattice magnetizations. An introduction of an external magnetic field is another way of perturbation of the highly degenerated triangular lattice antiferromagnetic system and, therefore, we believe that it would be interesting to perform a similar study of the combined effects of the selective dilution and the external field.   

\section*{Acknowledgments}
This work was supported by the Scientific Grant Agency of Ministry of Education of Slovak Republic (Grant No. 1/0234/12). The authors acknowledge the financial support by the ERDF EU (European Union European regional development fund) grant provided under the contract No. ITMS26220120005 (activity 3.2.).


\end{document}